\documentclass[showpacs,superscriptaddress,twocolumn,amsmath]{revtex4}
\usepackage{amssymb}
\usepackage{graphicx,color}
\usepackage{bm}
\usepackage[dvipdfm,bookmarks=true]{hyperref}
\newcommand{\ti}{\Tilde}

\newcommand{\nl}{\nonumber \\}
\newcommand{\nla}{\nl&\quad}

\newcommand{\up}{\uparrow}
\newcommand{\down}{\downarrow}

\newcommand{\be}{\begin{equation}}
\newcommand{\ee}{\end{equation}}
\newcommand{\bea}{\begin{eqnarray}}
\newcommand{\eea}{\end{eqnarray}}
\newcommand{\bsube}{\begin{subequations}}
\newcommand{\esube}{\end{subequations}}
\newcommand{\Eq}[1]{Eq.\,(\ref{#1})}

\newcommand{\dg}{\dagger}
\newcommand{\la}{\langle}
\newcommand{\ra}{\rangle}

\begin{document}
%\begin{CJK}{GBK}{song}
\draft

\topmargin=-40pt

\title{Nonequilibrium shot noise spectrum through a quantum dot in the Kondo
regime: \\ A master equation approach under self-consistent Born approximation}

\author{ Yu Liu}
\affiliation{State Key Laboratory for Superlattices and Microstructures,
         Institute of Semiconductors,
         Chinese Academy of Sciences, Beijing 100083, China}
\author{ Jinshuang Jin}
\email{jsjin@hznu.edu.cn}
\affiliation{Department of Physics, Hangzhou Normal University,
          Hangzhou 310036, China}
\author{Jun Li}
\affiliation{ Beijing Computational Science Research Center,
              Beijing 100084, China }
\author{ Xin-Qi Li}
\email{lixinqi@bnu.edu.cn}
\affiliation{Department of Physics, Beijing Normal University,
          Beijing 100875, China}
\affiliation{State Key Laboratory for Superlattices and Microstructures,
         Institute of Semiconductors,
         Chinese Academy of Sciences, Beijing 100083, China}
\author{YiJing Yan}
\affiliation{Department of Chemistry, Hong Kong University of Science
             and Technology, Kowloon, Hong Kong}

\begin{abstract}
We construct a number($n$)-resolved master equation (ME)
approach under self-consistent Born approximation (SCBA)
for noise spectrum calculation.
The formulation is essentially non-Markovian
and incorporates properly the interlay
of the multi-tunneling processes and many-body correlations.
We apply this approach to the challenging
nonequilibrium Kondo system and predict a profound
nonequilibrium Kondo signature in the shot noise spectrum.
The proposed $n$-SCBA-ME scheme goes completely beyond
the scope of the Born-Markovian master equation approach,
in the sense of being applicable to the shot noise of
transport under small bias voltage,
in non-Markovian regime, and with strong Coulomb correlations
as favorably demonstrated in the nonequilibrium Kondo system.
\end{abstract}

\date{\today}
%%03.65.Ta, 42.50.Lc, 02.30.Yy}
\maketitle

%%\section{Introduction}

Beyond the average current, shot noise (current fluctuations)
can provide deep insight
into the nature of transport mechanisms \cite{But00}.
In the past decade, most efforts have been devoted to
the zero- and low-frequency noise, including also the
full counting statistics \cite{Lev96}.
However, even more information is stored in the finite-frequency (FF)
current noise \cite{Mar01,Vish03,Nay06,Gab08,Hei09}.
For instance, the FF noise is sensitive to quantum statistics,
where a crossover between different statistics
can be revealed in the frequency domain.
Also, in the quantum regime, which is defined by frequencies
higher than the applied voltage or temperature,
the FF noise is a powerful tool
to probe the characteristic timescales of the system dynamics
associated with intrinsic excitations and interactions.

Among the various techniques for shot noise calculation
(including the counting statistics), the master equation approach,
particularly its number($n$)-resolved version
\cite{Gur96,Sch01,Jau05,Li05}, might be the most convenient one.
However, this technique is built largely on
the 2nd-order Born-Markovian master equation,
which limits thus its application only in zero- or low-frequency
noise, and under large bias voltage.
In this work, we will first extend the master equation approach
beyond these limits, making it highly non-Markovian and properly
account for the interplay of multiple tunneling and many-body correlations.
We then apply this new approach to the challenging
nonequilibrium Kondo system to calculate the FF noise spectrum,
where a profound Kondo resonance behavior will be revealed.

The nonequilibroum Kondo system, with the Anderson impurity
realized by transport through a small quantum dot (QD),
has been attracted intensive attention in the past two decades
\cite{Gor98,Kou98,Gla04,Ng88,Her91,MW92,MW9193,Ra94,Mar06,Gor08,Yan12}.
Compared to the {\it equilibrium} Kondo effect,
the {\it nonequilibrium} is characterized by
a finite chemical potential difference of the two leads.
As a result, the peak of the density of states (spectral function)
splits into two peaks pinned at each chemical potential.
The two peak structure is difficult to probe directly,
by the usual dc measurements.
Nevertheless, the shot noise can be a promising quantity
to reveal the nonequilibrium Kondo effect,
although much less is known about it.
We notice that results on low-frequency
noise measurements have only appeared very recently \cite{De09,Hei08},
while so far there are not yet reports on the FF noise measurements.
A couple of theoretical studies \cite{Ng97,Her98,Kon07,Moc11}, however,
revealed diverse signatures (Kondo anomalies) in the FF noise spectra,
such as an ``upturn" \cite{Ng97} or a spectral ``dip" \cite{Moc11}
appeared at frequencies $\pm eV/\hbar$ ($V$ is the bias voltage),
as well as the Kondo singularity (discontinuous slope)
at frequencies $\pm 2eV/\hbar$ in Ref.\ \cite{Her98},
or at $\pm eV/2\hbar$ in Ref.\ \cite{Moc11}.
Also, it was pointed out in Ref.\ \cite{Her98} that
the minimum (dip) developed at $\pm eV/\hbar$
is not relevant to the Kondo effect,
since in the noninteracting case the noise
has similar discontinuous slope at $\pm eV/\hbar$ as well.

%%\section{Formulation of SCBA-ME}
\vspace{0.2cm}

In general we describe a transport setup by
%%\be\label{H-T}
$ H = H_S(a_{\mu}^{\dg},a_{\mu})+ H_{\rm res} + H'$.
%%\ee
Here $H_S$ is the Hamiltonian of the central
{\it system} embedded between two leads,
with $a^{\dg}_{\mu}$ ($a_{\mu}$) the creation
(annihilation) operator of the state $|\mu\ra$.
More specifically, for a small and strongly interacting quantum dot,
with only a single level involved in transport
to realize an artificial Anderson impurity, we have
\be\label{Hs}
H_S = \sum_{\mu=\uparrow,\downarrow}
   \left(\epsilon_{\mu} a_{\mu}^{\dg}a_{\mu}
   + \frac{U}{2} n_{\mu}n_{\bar{\mu}}\right) .
\ee
In this model we use $\mu$ to label the spin-up (``$\uparrow$")
and spin-down (``$\downarrow$") states, and $\bar{\mu}$
corresponds to the opposite spin orientation.
$\epsilon_{\mu}$ is the spin-dependent (single) energy level,
and $U$ the on-site Coulomb repulsive energy
(with $n_{\mu}=a_{\mu}^{\dg}a_{\mu}$ the occupation number operator).
The other two Hamiltonians,
$H_{\rm res}$ and $H'$, describe the leads and their tunnel coupling
to the central system.
They are modeled by, respectively,
$H_{\rm res}=\sum_{\alpha=L,R}\sum_{k}\epsilon_{\alpha k}
b^{\dg}_{\alpha k}b_{\alpha k}$ and
$H'=\sum_{\alpha=L,R}\sum_{\mu k}(t_{\alpha\mu k}
a^{\dg}_{\mu}b_{\alpha k}+\rm{H.c.})$
with $b^{\dg}_{\alpha k}$ ($b_{\alpha k}$)
the creation (annihilation) operator of
electron in state $|k\ra$
of the left ($L$) and right ($R$) leads.

%%\subsection{SCBA-ME}
%%\vspace{0.2cm}

For the study of shot noise, the nonequilibrium Green's function
based calculation scheme is not efficient.
In contrast, an alternative one, say,
the particle-number($n$)-resolved master equation ($n$-ME)
plus the MacDonald's formula \cite{Mac62},
provides a much more convenient method for that purpose.
Also, the $n$-ME is extremely suitable
for studying the full counting statistics (FCS).
To our knowledge, the existing $n$-ME
scheme is only precise to the Born approximation (BA),
i.e., up to the 2nd-order expansion of the tunnel Hamiltonian
\cite{Gur96,Sch01,Jau05,Li05}.
Unfortunately, however, this type of master equation
cannot describe the small bias transport, since in this case
the multiple tunneling process between the system
and lead is heavily involved.
For similar reason, obviously, it {\it cannot} at all describe
the Kondo effect, which is actually a consequence of interplay
of the multiple tunneling and the many-body correlation.
Therefore, in order to study the shot noise
behavior through an interacting QD in the Kondo regime,
one has to include the effect of higher order
tunneling process in the master equation.
In a recent work \cite{LJL11}, going beyond the Born approximation,
an improved scheme under the self-consistent
Born approximation was proposed as follows:
\begin{align}\label{SCBA-ME}
  \dot\rho(t) &=-i{\cal L}\rho(t)
  - \sum_{\mu\sigma}\Big\{\big[a^{\bar\sigma}_\mu,
  {\cal A}^{(\sigma)}_{\mu\rho}(t)\big]
  +{\rm H.c.}      \Big\} .
\end{align}
Here we set the Planck constant $\hbar=1$ and will make further
convention in the following for a system of units
by setting $e=k_B=1$ for the electron charge and the Boltzmann constant.
In \Eq{SCBA-ME}, we define: $\sigma=+$ and $-$, $\bar{\sigma}=-\sigma$;
$a^+_{\mu}=a^{\dagger}_{\mu}$, and $a^-_{\mu}=a_{\mu}$.
%In \Eq{SCBA-ME}, the Liouvillian superoperator is defined as
%${\cal L}\rho=[H_S,\rho]$.
%While in this work only the Anderson impurity Hamiltonian
%is considered, we would like to emphasize that
%the SCBA-ME formulation is applicable to arbitrary $H_S$.
Also, the superoperators in \Eq{SCBA-ME} read
${\cal L}\rho=[H_S,\rho]$ and
$ {\cal A}^{(\sigma)}_{\mu\rho}(t)
= \sum_{\alpha=L,R} {\cal A}^{(\sigma)}_{\alpha\mu\rho}(t)$
while
%%\begin{align}\label{Arho-SCBA}
$ {\cal A}^{(\sigma)}_{\alpha\mu\rho}(t)
 =\sum_\nu\int^t_0 d\tau C^{(\sigma)}_{\alpha\mu\nu}(t-\tau)
\left\{{\cal U}(t,\tau)[a^{\sigma}_\nu\rho(\tau)]\right\} $.
$ C^{(\sigma)}_{\alpha\mu\nu}(t-\tau)$
is the reservoir correlation function
(see Appendix A for more details).
Very importantly,
${\cal U}(t,\tau)$ is an {\it effective}
propagator under the spirit of SCBA,
which considerably generalizes the $H_S$-defined {\it free}
propagator ${\cal G}(t,\tau)=e^{-i{\cal L}(t-\tau)}$
in the 2nd-order Born master equation.
The SCBA is implemented by defining
$\ti\rho_j(t)\equiv {\cal U}(t,\tau)[a^{\sigma}_\nu\rho(\tau)]$
(here and in the following we use ``$j$" to denote the double
indices $(\nu,\sigma)$ for the sake of brevity),
and closing \Eq{SCBA-ME}
via an equation-of-motion (EOM) for this auxiliary object:
\begin{align}\label{tirhoj}
  \dot{\ti\rho}_j(t) = -i{\cal L}\ti\rho_j(t)
-\int^t_{\tau}dt'  \Sigma^{(A)}_2(t-t')\ti\rho_j(t').
\end{align}
In this equation the 2nd-order self-energy superoperator,
$\Sigma^{(A)}_2(t-t')$, differs from the usual one
since it involves {\it anticommutators}, but not the {\it commutators}
in the 2nd-order master equation
(See Appendix A for an explicit expression).

%%\subsection{$``n"$-resolved master equation (SCBA-nME)}

Now we proceed further to construct the particle number (``$n$")
resolved SCBA-ME. To be specific, consider the reduced system
state $\rho^{(n)}$, conditioned on the electron number
arrived to the {\it right} lead, which satisfies
\begin{align}\label{nME-scba}
\dot{\rho}^{(n)}
   & =  -i {\cal L}\rho^{(n)} -  \sum_{\mu}
%\nl&\quad
   \Big\{\big [a_{\mu}^{\dg} {\cal A}_{\mu\ti{\rho}^{(n)}}^{(-)}
 %\nl&\quad
   +a_{\mu} {\cal A}_{\mu\ti{\rho}^{(n)}}^{(+)}
 %\nl &\quad
    -  {\cal A}_{L\mu\ti{\rho}^{(n)}}^{(-)}a_{\mu}^{\dg}
  \nl
&  -  {\cal A}_{L\mu\ti{\rho}^{(n)}}^{(+)}a_{\mu}
 %\nl&\quad
   -  {\cal A}_{R\mu\ti{\rho}^{(n-1)}}^{(-)}a_{\mu}^{\dg}
 %\nl&\quad
    -  {\cal A}_{R\mu\ti{\rho}^{(n+1)}}^{(+)}a_{\mu}\big]
 %\nl&\qquad
 +{\rm H.c.} \Big\}  .
%\nl & \equiv -i {\cal L}\rho^{(n)} + {\cal R}_{\ti{\rho}^{(n)}}
%   + {\cal R}^{(+)}_{\ti{\rho}^{(n+1)}}
%   + {\cal R}^{(-)}_{\ti{\rho}^{(n-1)}}  .
\end{align}
%In this result we introduced
%\begin{align}\label{Arhon-SCBA}
Here $ {\cal A}^{(\sigma)}_{\alpha\mu\ti{\rho}^{(n)}}(t)
=\sum_\nu\int^t_0 d\tau C^{(\sigma)}_{\alpha\mu\nu}(t-\tau)
[ \ti{\rho}_j^{(n)}(t,\tau)] $, while the summation over $\nu$
makes sense in regard to the abbreviation of $j=\{\nu,\sigma\}$.
%%\left\{\widetilde{\cal U}(n,t,\tau)[a^{\sigma}_\nu\rho^{(n)}(\tau)]\right\}
%\end{align}
%$\sum_\nu\int^t_0 d\tau C^{(\sigma)}_{\alpha\mu\nu}(t-\tau)
%\left\{\widetilde{\cal U}(n,t,\tau)[\ti\rho^{(n)}_j(\tau)]\right\}.$
%%
In particular, $\ti{\rho}_j^{(n)}(t,\tau)$
is the $n$-dependent version of the quantity
$\ti{\rho}_j(t,\tau)={\cal U}(t,\tau)[a^{\sigma}_{\nu}\rho(\tau)]$,
satisfying an EOM according to \Eq{tirhoj}:
%Note that although the propagator $\widetilde{\cal U}(n,t,\tau)$ in \Eq{Arhon-SCBA}
%is the second-order expansion, it is different from ${\cal U}(t,\tau)$
%in \Eq{Arho-SCBA}. The relation of
%${\cal A}^{(\sigma)}_{\alpha\mu\rho}(t)=\sum_n \ti{\cal A}^{(\sigma)}_{\alpha\mu\rho^{(n)}}(t)$
%or equivalently, ${\cal U}(t,\tau)\ti\rho_j(\tau)
%=\sum_n\widetilde{\cal U}(n,t,\tau)\ti\rho^{(n)}_j(\tau)$
%should be satisfied to recover SCBA-ME in \Eq{QME-SCBA}. Explicitly,
%the time evolution of $\ti\rho^{(n)}_j(t)\equiv \widetilde{\cal U}(n,t,t_0)\ti\rho^{(n)}_j(t_0)$ satisfies
\begin{align}\label{nME-2nd}
\dot{\ti\rho}^{(n)}_{j}
   &=  -i {\cal L}\ti\rho^{(n)}_{j} -  \sum_{\mu}
%\nl&\quad
   \Big\{\big [a_{\mu}^{\dg} A_{\mu\tilde{\rho}^{(n)}_{j}}^{(-)}
 %\nl&\quad
   +a_{\mu} A_{\mu\tilde{\rho}^{(n)}_{j}}^{(+)}
 %\nl &\quad
    +  A_{L\mu\tilde{\rho}^{(n)}_{j}}^{(-)}a_{\mu}^{\dg}
\nl& %\quad
   +  A_{L\mu\tilde{\rho}^{(n)}_{j}}^{(+)}a_{\mu}
 %\nl&\quad
   +  A_{R\mu\tilde{\rho}^{(n-1)}_{j}}^{(-)}a_{\mu}^{\dg}
 %\nl&\quad
    +  A_{R\mu\tilde{\rho}^{(n+1)}_{j}}^{(+)}a_{\mu}\big]
 %\nl&\qquad
 +{\rm H.c.} \Big\} .
% \nl&
% \equiv -i {\cal L}\ti\rho^{(n)}_{j}
% + R_{\ti\rho^{(n)}_{j}} + R^{(+)}_{\ti\rho^{(n+1)}_{j}}
% + R^{(-)}_{\ti\rho^{(n-1)}_{j}}
\end{align}
In this equation we introduced
$ A^{(\sigma')}_{\alpha\mu\ti\rho^{(n)}_{j}}(t)
 =\sum_{\nu'}\int^t_\tau dt' C^{(\sigma')}_{\alpha\mu\nu'}(t-t')
\left\{ e^{-i{\cal L}(t-t')}[a^{\sigma'}_{\nu'}
\tilde{\rho}_j^{(n)}(t')]\right\}$.

%% ========================

The structure of \Eq{nME-scba} follows the same idea of constructing
the 2nd-order $n$-resolved master equation \cite{Gur96,Li05},
which is essentially equivalent to
the {\it counting-field} approach \cite{Sch01,Jau05}.
Following \cite{Li05},
we split the Hilbert space of the electron reservoirs
into a set of subspaces, each labeled by $n$.
Then we do the average (trace) over each subspace
and define the corresponding {\it reduced} quantities as
$\rho^{(n)}(t)$ and $\tilde{\rho}^{(n)}_j(t,\tau)$.
In \Eq{nME-scba}, moreover,
the appearance of $\tilde{\rho}^{(n\pm 1)}_j(t,\tau)$
is owing to a more tunneling event (forward/backword)
involved in the process of the corresponding terms.
These considerations also lead to the $n$-dependence structure of
\Eq{nME-2nd}, the EOM of the auxiliary quantity $\tilde{\rho}^{(n)}_j$.

%%% ================================
%\vspace{1cm}
%一些说明：
%1)  eq (3)中， “$n$”是$(0,t)$时间段的tunnel电子数。
%与之相对应(conditioned on it)，我们引入了两个
%$n$-dependent quantity: one is $\rho^{(n)}$;
%another is $\tilde{\rho}^{(n)}_j$.
%特别注意：$\tilde{\rho}^{(n)}_j(t,\tau)$中的"$n$",
%不是$(\tau,t)$时间段积累的"$n$",也不意味其中隐含的
%$\rho(\tau)$ 对应为$\rho^{(n)}(\tau)$ [以前是这样处理的]。\\
%2)  (关键)问题：$\tilde{\rho}^{(n)}_j$中 $n$ 的变化由什么“驱动”？
%答：任何 $n$ 到 $n\pm 1$的变化，都只能由一对jump (tunnel)算符引起；
%具体如eq (4) 所示。   \\
%3)  按tunnel electron number对电极的 “态空间”分类，理解方式为： 。。。
%\\
%4)  关于后面eq(8)的理解：由于是“从稳态出发开始counting”，所以
%$\rho(\tau)$始终用稳态代替，且“作为初条件进入eq(4)”的求解。
%..// 小心：$\sum_n n \ti{\rho}_j^{(n)}(t,\tau)\equiv \ti{N}_j(t)$,
%当$t$趋于$\tau$时，$\ti{N}_j$是否为零？

%% \clearpage  ========================================
% \subsection{Noise spectrum}
\vspace{0.2cm}

The noise spectrum, $S_I(\omega)$, is the Fourier transform
of the current correlation function
$S_I(t)=\la I(t) I(0)\ra_{ss}$ defined in the steady state.
Very conveniently, within the framework of the $n$-ME,
one can calculate $S_I(\omega)$
by using the MacDonald's formula \cite{Mac62}:
$S_I(\omega)=2\omega\int^{\infty}_{0}dt \sin(\omega t)
\frac{d}{dt}\langle n^{2}(t)\rangle$,
where
$\langle n^{2}(t)\rangle = \sum_{n}n^{2}P(n,t)= {\rm Tr}
\sum_{n} n^{2} \rho^{(n)}(t)$, and the $n$-counting
starts with the steady state ($\bar{\rho}$).
Based on \Eq{nME-scba}, one can express
$\frac{d}{dt} \langle n^{2}(t)\rangle$
in terms of ${\cal A}^{(\sigma)}_{R\mu\bar{\rho}}(t)$
and ${\cal A}^{(\sigma)}_{R\mu\ti{N}}(t)$.
The former has been introduced in \Eq{SCBA-ME},
needing only to replace $\rho(\tau)$ by $\bar{\rho}$.
The latter reads $ {\cal A}^{(\sigma)}_{R\mu\ti{N}}(t)
=\sum_\nu\int^t_0 d\tau C^{(\sigma)}_{R\mu\nu}(t-\tau)
[ \ti{N}_j(t,\tau)]$, where
$\ti{N}_j(t,\tau)=\sum_n n \ti{\rho}_j^{(n)}(t,\tau)$.
Then, the MacDonald's formula becomes
\begin{align}\label{Sw-scba}
S_I(\omega)&= 2\omega {\rm Im}\sum_{\mu}{\rm Tr}\Big\{ 2 \big[
 {\cal A}_{R\mu \ti{N}}^{(-)}(\omega)a_{\mu}^{\dg}
 %\nl&\quad
    - {\cal A}_{R\mu \ti{N}}^{(+)}(\omega)a_{\mu}\big]
 \nl&\qquad\qquad
  %\nl&\qquad
  +  \big[
  {\cal A}_{R\mu\bar{\rho}}^{(-)}(\omega)a_{\mu}^{\dg}
 %\nl&\quad
    + {\cal A}_{R\mu\bar{\rho}}^{(+)}(\omega)a_{\mu}\big]
  \Big\} .
\end{align}
This result is obtained after Laplace transforming
${\cal A}^{(\sigma)}_{R\mu\bar{\rho}}(t)$
and ${\cal A}^{(\sigma)}_{R\mu\ti{N}}(t)$. More explicitly,
\be
%\begin{align}\label{SW-1}
{\cal A}^{(\sigma)}_{R\mu\bar{\rho}}(\omega)
= \sum_\nu\int^\infty_{-\infty} \frac{d\omega'}{2\pi}
\Gamma^{(\sigma)}_{R\mu\nu}(\omega')
{\cal U}(\omega+\sigma\omega')[a^{\sigma}_\nu\bar{\rho}(\omega)], \nonumber
%\end{align}
\ee
where the Laplace transformation of the steady state
reads $\bar{\rho}(\omega)=i\bar{\rho}/\omega$,
and the propagator ${\cal U}$ in frequency domain
is defined through \Eq{tirhoj}.
On the other hand, ${\cal A}^{(\sigma)}_{R\mu \ti{N}}(\omega)$ reads
%\begin{align}\label{SW-2}
\be
{\cal A}^{(\sigma)}_{R\mu \ti{N}}(\omega)
= \sum_\nu\int^\infty_{-\infty} \frac{d\omega'}{2\pi}
\Gamma^{(\sigma)}_{R\mu\nu}(\omega')
\ti{\cal U}(\omega+\sigma\omega')[a^{\sigma}_{\nu}N(\omega)].  \nonumber
%% {\cal U}(\omega+\sigma\omega')[a^{\sigma}_\nu N(\omega)],
%\end{align}
\ee
In deriving this result, we introduced an additional propagator
through $\ti{N}_j(t,\tau)=\ti{\cal U}(t-\tau)\ti{N}_j(\tau)$,
with $\ti{N}_{j}(\tau)=a^{\sigma}_{\nu} N(\tau)$ as the initial condition
which is defined by $N(\tau)=\sum_n n \rho^{(n)}(\tau)$.
$\ti{\cal U}(\omega)$ and $N(\omega)$ can be obtained via
Laplace transforming the following EOMs.
%%======================
{\it (i)} For $N(\omega)$,
based on the $n$-SCBA-ME we obtain
\begin{align}\label{N-t}
  \dot{N}(t) &=-i{\cal L}N(t)
  - \sum_{\mu\sigma}\Big\{\big[a^{\bar\sigma}_\mu,
  {\cal A}^{(\sigma)}_{\mu N}(t)\big]
  +{\rm H.c.}  \Big\}  \nl
& ~~ + \sum_{\mu} \Big\{\big[ {\cal A}_{R\mu\bar{\rho}}^{(-)}a_{\mu}^{\dg}
    - {\cal A}_{R\mu\bar{\rho}}^{(+)}a_{\mu}\big] +{\rm H.c.} \Big\}  .
\end{align}
%=============================
{\it (ii)}
For $\ti{\cal U}(\omega)$, from \Eq{nME-2nd} we have
\begin{align}\label{Nj-2nd}
& \dot{\ti N}_{j}(t)
   =  -i {\cal L}\ti N_{j}(t)
   - \int^t_{\tau} dt' {\Sigma}^{(A)}_2(t-t')\ti N_{j}(t')
\nl& ~
 - \sum_{\mu} \Big\{\big [A_{R\mu\ti{\rho}_{j}}^{(-)}(t)a_{\mu}^{\dg}
-A_{R\mu\ti{\rho}_{j}}^{(+)} (t)a_{\mu}\big]
 %\nl&\qquad
 +{\rm H.c.} \Big\}.
\end{align}
The self-energy superoperator ${\Sigma}^{(A)}_2(t-t')$
is referred to \Eq{Acmt} in Appendix A for its definition.
Similar as introduced in \Eq{nME-2nd}, we defined here
%$ A^{(\pm)}_{R\mu\rho_{j}}(t)
% =\sum_\nu\int^t_{\tau} dt' C^{(\pm)}_{R\mu\nu}(t-t')
%\left\{ e^{-i{\cal L}(t-t')}[a^{\pm}_{\nu}\rho(t')]\right\}$.
$ A^{(\sigma')}_{R\mu\ti\rho_{j}}(t)
 =\sum_{\nu'}\int^t_\tau dt' C^{(\sigma')}_{R\mu\nu'}(t-t')
\left\{ e^{-i{\cal L}(t-t')}[a^{\sigma'}_{\nu'}
\tilde{\rho}_j(t')]\right\}$.
%%
%Notice that when solving $\ti{N}_{j}(\omega)$
%in frequency domain, we will encounter $\rho(\omega)$.
%This quantity, however, is directly given
%by Laplace transforming the SCBA-ME.

For the convenience of application, we would like to summarize
the solving protocol in a more transparent way as follows.
First, solve ${\cal U}(\omega)$ from \Eq{tirhoj}
and obtain $\rho(\omega)$ from \Eq{SCBA-ME};
then, extract $\tilde{\cal U}(\omega)$ from \Eq{Nj-2nd}
and $N(\omega)$ from \Eq{N-t}.
With the help of ${\cal U}(\omega)$,
$\tilde{\cal U}(\omega)$ and $N(\omega)$, one can
straightforwardly calculate the noise spectrum of \Eq{Sw-scba}.

%%====================================================
\vspace{0.2cm}

Now we return to the Anderson impurity model. Simply,
there are four states involved in transport:
$|0\ra$, $|\up\ra$, $|\down\ra$ and $|d\ra$, which correspond to
the empty, spin-up, spin-down and double occupancy states, respectively.
With respect to these states, the reservoir correlation
function $C^{(\pm)}_{\alpha\mu\nu}$ is diagonal, i.e.,
$C^{(\pm)}_{\alpha\mu\nu}(t)=\delta_{\mu\nu}C^{(\pm)}_{\alpha\mu}(t)$
and
$\Gamma^{(\pm)}_{\alpha\mu\nu}=\Gamma^{(\pm)}_{\alpha\mu}\delta_{\mu\nu}$.
Moreover, using these basis states, we can reexpress the electron operator
in terms of the projection operator form,
$a^\dg_\mu=|\mu\ra\la 0|+(-1)^{\mu}|d\ra\la \bar \mu|$,
where the convention $(-1)^\up=1$ and $(-1)^\down=-1$ is assumed.
Since the shot noise spectrum is defined on the steady-state current
fluctuations, we need first a solution of the steady state ($\bar{\rho}$).
In steady state, one can express the key operator in \Eq{SCBA-ME} as
%\begin{align}\label{Arhost-Ad}
${\cal A}^{(\pm)}_{\alpha\mu\bar\rho}
=\int^\infty_{-\infty}\frac{d\omega}{2\pi}\,
\Gamma^{(\pm)}_{\alpha\mu}(\omega)
{\cal U}(\pm\omega)[a^{\pm}_\mu\bar\rho]$.
%\end{align}
Straightforwardly, after some algebra, we obtain \cite{LJL11}
%\bsube
\begin{align}\label{U-kondo}
{\cal U}(\omega)[a^{\dg}_\mu\bar\rho]
&=\left[\lambda^+_\mu(\omega)|\mu\ra\la 0|
+\kappa^+_\mu(\omega)(-1)^{\mu}|d\ra\la \bar \mu|\right] , \nonumber
\\
{\cal U}(-\omega)[a_\mu\bar\rho]&=\left[\lambda^-_\mu(\omega)|0\ra\la \mu|
+\kappa^-_\mu(\omega)(-1)^{\mu}|\bar \mu\ra\la d|\right] .
\end{align}
In terms of the matrix elements of $\bar{\rho}$,
the specific expressions of
$\lambda^{\pm}_\mu(\omega)$ and $\kappa^{\pm}_\mu(\omega)$
are given in Appendix B.
Substituting ${\cal A}^{(\pm)}_{\alpha\mu\bar\rho}$,
with the result of \Eq{U-kondo}, into \Eq{SCBA-ME}
one can first obtain the steady state.
Then, following the solving protocol outlined above,
the noise spectrum can be carried out.
%the subsequent state evolution after the {\it counting disturbance}
%on the steady state, which are characterized by
%$\rho(\omega)$, $N(\omega)$, ${\cal U}(\omega)$ and $\tilde{\cal U}(\omega)$,
%and finally to

%% ===============================================
\vspace{0.2cm}

\begin{figure}
\includegraphics[scale=0.4]{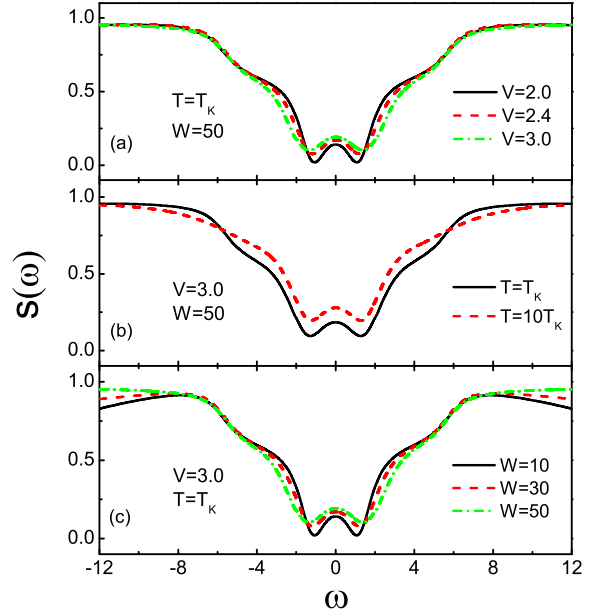}
\caption{
Shot noise spectrum in the Kondo regime, by varying the bias voltage (a),
the temperature (b) and the bandwidth ($W$) of the reservoirs (c).
We assume $\hbar=e=k_B=1$ and use an arbitrary unit
of energy in this model simulation,
with parameters as $\Gamma_L=\Gamma_R=\Gamma=0.5$,
$\epsilon_{\uparrow}=\epsilon_{\downarrow}=\epsilon=-2$, and $U=6$.
The bias voltage is defined as usual by $\mu_L=-\mu_R=V/2$.
The Kondo temperature is given by
$T_K=\frac{U}{2\pi}\sqrt{\frac{-2U\Gamma}{\epsilon(U+\epsilon)}}
\exp[\frac{\pi\epsilon(U+\epsilon)}{2U\Gamma}]$,
having a value of $T_{K}=0.144$ for the given parameters.    }
\end{figure}

%we assume the model parameters as: $E_0=-2$,
%$\mu_L=-\mu_R=1.44$ (thus $eV=2.88$), $U=6$,
%and the Kondo temperature $T_K=0.144$.
%%
In Fig.\ 1 we display the symmetrized shot noise spectrum
in Kondo regime (the numerical results are presented
with the use of $\hbar=e=k_B=1$).
First of all, we notice a remarkable {\it dip} behavior (Kondo signature)
in the noise spectrum at the frequencies $\omega=\pm V/2$,
as particularly demonstrated in Fig.\ 1(a) by altering the voltages.
We attribute this behavior to the emergence
of the Kondo resonance levels (KRLs)
induced at the Fermi surfaces,
i.e., at $\mu_L=V/2$ and $\mu_R=-V/2$.
In steady state transport, it is well known that
the KRLs are clearly reflected in the spectral function, i.e.,
the effective density of states (DOS) of the Anderson impurity.
In terms of the master equation (see Appendix B),
the KRLs structure is hidden in the self-energy terms,
which characterize the tunneling process
and define the transport current.
Similarly, the noise spectrum is essentially affected,
particularly in the Kondo regime,
by the self-energy process in frequency domain
based on the same master equation.
This explains the emergence of the spectral dip appearing
at the same KRLs (i.e., at $\omega=\pm V/2$).

However, we would like to remark that the dip behavior is also a
consequence of highly non-Markovian treatment of the current correlations.
We have checked that, using the quantum-jump technique \cite{Mil99}
or the quantum regression theorem \cite{GZ00},
this behavior cannot be recovered, even the evolution during $(0,t)$
is treated as non-Markovian based on \Eq{SCBA-ME}.
The point is that the definition of the current in the correlation function
$\la I(t) I(0) \ra$, in the non-Markovian case,
cannot be {\it independent} of the propagation during the time interval $(0,t)$,
because of the non-Markovian {\it memory} effect.
In contrast, based on the $n$-SCBA-ME, the MacDonald
formula correctly accounts for the correlation between
the current and the memory effect during $(0,t)$,
by employing the number($n$)-counting technique.

Alternatively, as a heuristic picture, one may imagine to include
the KRLs as basis states in propagating $\rho(t)$,
which is implied in the current correlation function.
In usual case, when the level spacing is larger than its broadening,
the diagonal elements of the density matrix decouple to
the evolution of the off-diagonal elements.
However, in the Kondo system, the diagonal and off-diagonal elements
are coupled to each other, through the complicated self-energy processes.
This feature would bring the {\it coherence evolution}
described by the off-diagonal elements, with characteristic
energies of the KRLs and their difference, into the diagonal elements
which contribute directly to the the second current measurement
in the correlation function $\la I(t) I(0) \ra$.
Then, one may expect three coherence energies, $\pm V/2$ and $V$,
to participate in the noise spectrum.
Indeed, the dip emerged in Fig.\ 1 reveals the {\it coherence}-induced
oscillation at the frequencies $\pm V/2$,
while the other one at the higher frequency $V$
(observed in Ref.\ \cite{Moc11} in the case of infinite $U$)
is smeared in our finite $U$ system by the rising noise
with frequency.

Physically, the current fluctuation spectrum corresponds to electron
transfer between the dot and leads, accompanied by the energy
($\omega$) absorption/emission of detection.
Therefore, as the frequency ($\omega$) matches the energy difference
between the dot level and the Fermi surface of the lead, certain
``singularity" associated with the Fermi function
at the Fermi surface is expected to emerge in the spectrum.
This is reflected in Fig.\ 1(a) by the staircase behavior.
This ``singularity", however, has been smoothed by the finite
temperature effect (see Fig.\ 1(b) for further illustration).
%Also, with the increase of temperature, as shown in Fig.\ 1(b),
%the Kondo resonance (dip) behavior becomes weaker,
%and will disappear eventually.
%%
In Fig.\ 1(c) we display the bandwidth effect.
For finite (narrow) bandwidth, the spectrum would
diminish at high frequencies (when much higher than the bandwidth),
since in this case the electron transfer channel
associated with the $\omega$-emission/absorption is switched off.
In the low frequency regime, on the other hand, we find that
the narrowing bandwidth would shift the Kondo dip to lower frequency.
This feature indicates that the Kondo peak pinned at the chemical
potential is only a result in the wide band limit.
For finite (especially narrow) bandwidths, it may need further
work to determine the location of the Kondo peaks.

\vspace{0.1cm}

To summarize, we have applied a new shot noise scheme
to the nonequilibrium Kondo system,
for finite $U$ and arbitrary bandwidths.
The scheme is based on a generalized
number($n$)-resolved master equation
under self-consistent Born approximation,
which considerably goes beyond the scope of the usual
2nd-order Born master equation.
This treatment allows us to predict
a profound nonequilibrium Kondo signature in shot noise
at frequencies associated with the chemical potentials.
We anticipate a wide range of applications of the proposed
approach to shot noise studies, as well as future work
to clarify the diverse
Kondo signatures in noise spectrum \cite{Ng97,Her98,Kon07,Moc11}.

%% ===============================================
\vspace{0.5cm}
%\begin{acknowledgements}
{\it Acknowledgements.}---
This work was supported by the NNSF of China,
%% under grants  No.\ 101202101 \& 10874176,
the Major State Basic Research Project of China
under grants 2011CB808502 \& 2012CB932704,
and the Fundamental Research Funds for the Central Universities of China.
J.J. was also supported by the Program for Excellent Young Teachers
in Hangzhou Normal University and by the NSFC under No.11274085.
%\end{acknowledgements}

\clearpage
\appendix

\section{Some Particulars in the SCBA-ME Approach}

\subsection{Reservoir Spectral Density Function}

The key operators in \Eq{SCBA-ME} read
$ {\cal A}^{(\sigma)}_{\mu\rho}(t)
= \sum_{\alpha=L,R} {\cal A}^{(\sigma)}_{\alpha\mu\rho}(t)$,
and $ {\cal A}^{(\sigma)}_{\alpha\mu\rho}(t)
 =\sum_\nu\int^t_0 d\tau C^{(\sigma)}_{\alpha\mu\nu}(t-\tau)
\left\{{\cal U}(t,\tau)[a^{\sigma}_\nu\rho(\tau)]\right\}   $.
$ C^{(\sigma)}_{\alpha\mu\nu}(t-\tau)$ are the correlation
functions of the reservoir electrons (in local equilibrium),
being defined as
\begin{align}
C^{(\sigma)}_{\alpha\mu\nu}(t-\tau)
= \la f^{(\sigma)}_{\alpha\mu}(t)
f^{(\bar\sigma)}_{\alpha \nu}(\tau) \ra_{\rm B}.
\end{align}
Here, $f^{(+)}_{\alpha\mu}(t)=f^{\dagger}_{\alpha\mu}(t)$
and $f^{(-)}_{\alpha\mu}(t)=f_{\alpha\mu}(t)$,
resulting from rewriting the tunneling Hamiltonian
$H'=\sum_{\alpha=L,R}\sum_{\mu k}(t_{\alpha\mu k}
a^{\dg}_{\mu}b_{\alpha\mu k}+\rm{H.c.})
= \sum_{\alpha=L,R}\sum_{\mu} \left( a^{\dg}_{\mu} f_{\alpha\mu}
       + \rm{H.c.}\right) $,
by introducing
$f_{\alpha\mu} = \sum_{k} t_{\alpha\mu k}b_{\alpha\mu k}$.
The time dependence of the operators in
$ C^{(\sigma)}_{\alpha\mu\nu}(t-\tau)$
originates from using the interaction picture
with respect to the reservoir Hamiltonian,
while the average $\la \cdots \ra_B$ is over the reservoir states.
Moreover, we introduce the Fourier transform of
$ C^{(\sigma)}_{\alpha\mu\nu}(t-\tau)$ through
\begin{align}
C^{(\pm)}_{\alpha\mu\nu}(t-\tau)=\int^\infty_{-\infty}
\frac{d\omega}{2\pi}
e^{\pm i\omega (t-\tau)}\Gamma^{(\pm)}_{\alpha\mu\nu}(\omega).
\end{align}
Accordingly, we have
$\Gamma^{(+)}_{\alpha\mu\nu}(\omega)
=\Gamma_{\alpha\nu\mu}(\omega)n^{(+)}_{\alpha}(\omega)$
and $\Gamma^{(-)}_{\alpha\mu\nu}(\omega)
=\Gamma_{\alpha\mu\nu}(\omega)n^{(-)}_{\alpha}(\omega)$,
where $\Gamma_{\alpha\mu\nu}(\omega)
=2\pi\sum_{k}t_{\alpha\mu k}t^\ast_{\alpha\nu k}\delta(\omega-\epsilon_k)$
is the spectral density function of the reservoir ($\alpha$),
$n^{(+)}_{\alpha}(\omega)$ denotes the Fermi function $n_{\alpha}(\omega)$,
and $n^{(-)}_{\alpha}(\omega)=1-n_{\alpha}(\omega)$ is introduced for brevity.
Alternatively, we may introduce as well the Laplace transform
of $ C^{(\sigma)}_{\alpha\mu\nu}(t-\tau)$, denoting by
$C^{(\sigma)}_{\alpha\mu \nu}(\omega)$,
which is related with $\Gamma^{(\pm)}_{\alpha\mu \nu}(\omega)$
through the well known dispersive relation:
\begin{align}\label{FDT2}
C^{(\pm)}_{\alpha\mu \nu}(\omega)
&=\int^\infty_{-\infty}\frac{d\omega'}{2\pi}
\frac{i}{\omega\pm\omega'+i0^+}\Gamma^{(\pm)}_{\alpha\mu \nu}(\omega').
\end{align}

In this work, for the reservoir spectral density function,
we assume a Lorentzian form as
\be\label{Gammaw}
\Gamma_{\alpha\mu \nu}(\omega)=
\frac{\Gamma_{\alpha\mu \nu} W^2_\alpha}{(\omega-\mu_\alpha)^2+W^2_\alpha} .
\ee
In some sense, this assumption corresponds to a half-occupied band
for each lead, which peaks the Lorentzian center
at the chemical potential $\mu_\alpha$.
$W_\alpha$ characterizes the bandwidth of the $\alpha$th lead.
Obviously, the usual constant spectral density function is
recovered from \Eq{Gammaw} in the limit $W_\alpha\rightarrow\infty$,
yielding $\Gamma_{\alpha\mu \nu}(\omega)=\Gamma_{\alpha\mu \nu}$.
Corresponding to the above Lorentzian spectral density function,
straightforwardly, we obtain
\begin{align}
C^{(\pm)}_{\alpha\mu \nu}(\omega)
&=\frac{1}{2}\left[\Gamma^{(\pm)}_{\alpha\mu \nu}(\mp \omega)
 +i\Lambda^{(\pm)}_{\alpha\mu \nu}(\mp \omega)\right].
\end{align}
The imaginary part, through the dispersive relation,
is associated with the real one as
\begin{align}
& \Lambda^{(\pm)}_{\alpha\mu \nu} (\omega)
={\cal P}\int^\infty_{-\infty}\frac{d\omega'}{2\pi}
\frac{1}{\omega\pm\omega'}\Gamma^{(\pm)}_{\alpha\mu \nu}(\omega)
\nl&=\frac{\Gamma_{\alpha\mu\nu}}{\pi}
\Bigg\{{\rm Re}\left[\Psi\left(\frac{1}{2}
+i\frac{\beta(\omega-\mu_\alpha)}{2\pi}\right)\right]
\nla
-\Psi\left(\frac{1}{2}+\frac{\beta W_\alpha}{2\pi}\right)
\mp\pi\frac{\omega-\mu_\alpha}{W_\alpha}\Bigg\},
\end{align}
where ${\cal P}$ stands for the principle value
and $\Psi(x)$ is the digamma function.

\subsection{Anomalous Self-Energy Superoperator}
%% {Effective Propagation in Self-Energy Process}

The central idea of the SCBA-ME scheme is
replacing the {\it free} propagator
in the 2nd-order master equation,
${\cal G}(t,\tau)=e^{-i{\cal L}(t-\tau)}$,
by an effective one, ${\cal U}(t,\tau)$ under the SCBA spirit.
By introducing
$\ti\rho_j(t)= {\cal U}(t,\tau)[a^{\sigma}_\nu\rho(\tau)]$,
we obtain \Eq{tirhoj}, the EOM of this auxiliary object.
In \Eq{tirhoj}, the 2nd-order self-energy superoperator,
$\Sigma^{(A)}_2(t-t')$, is worth receiving some special attention.
As labeled by the superscript ``$(A)$", an {\it anticommutator},
instead of the usual {\it commutator}, is involved there.
That is, the self-energy superoperator has the following form:
\begin{align}\label{Acmt}
\int^t_{\tau}dt' \Sigma^{(A)}_2(t-t')
& \ti\rho_j(t')= \sum_{\mu} \Big[\big\{a_\mu,A^{(+ )}_{\mu\ti\rho_j}\big\}
+\big\{a^\dg_\mu,A^{(-)}_{\mu\ti\rho_j}\big\}   \nl
& +\big\{a^\dg_\mu,A^{(+ )\dg}_{\mu\ti\rho_j}\big\}
+\big\{a_\mu,A^{(- )\dg}_{\mu\ti\rho_j}  \big\} \Big] ,
\end{align}
where $A^{(\pm)}_{\mu\ti\rho_j}$ is defined as
$ A^{(\sigma')}_{\mu\ti\rho_j}=\sum_{\alpha=L,R}\sum_{\nu'}
 \int^t_\tau dt' C^{(\sigma')}_{\alpha\mu\nu'}(t-t')
\left\{ e^{-i{\cal L}(t-t')}[a^{\sigma'}_{\nu'}
\tilde{\rho}_j(t')]\right\}$.
%%
%is the simpler version
%of ${\cal A}^{(\pm)}_{\mu\rho}$ in \Eq{SCBA-ME},
%by replacing ${\cal U}(t,t')$
%with the {\it free} propagator ${\cal G}(t,t')$.
%%
We remark that the {\it anticommutative} brackets appeared in \Eq{Acmt}
indicate that the propagation of $\ti\rho_j(t)$
does not satisfy the usual 2nd-order master equation.
This actually violates the so-called {\it quantum regression theorem}.

\subsection{Steady-State Current}

Similar to the usual 2nd-order master equation approach,
the current through the $\alpha$th lead reads
\begin{align}
   I_{\alpha}(t)=\frac{2e}{\hbar}\sum_\mu {\rm Re}
\left\{ {\rm Tr}\big[ {\cal A}^{(+)}_{\alpha\mu\rho}(t)a_\mu
  - {\cal A}^{(-)}_{\alpha\mu\rho}(t)a^\dg_\mu\big] \right\} .
\end{align}
Moreover, the steady state together with its associated current
can be obtained easily as follows.
Consider the integral $\int^t_0 d\tau [\cdots]\rho(\tau)$
in ${\cal A}^{(\pm)}_{\alpha\mu\rho}(t)$.
Since physically, the correlation function
$C^{(\pm)}_{\alpha\mu\nu}(t-\tau)$ in the integrand
is nonzero only on {\it finite} timescale,
we can replace $\rho(\tau)$
in the integrand by the steady state $\bar{\rho}$,
in the long time limit ($t\rightarrow\infty$).
After this replacement, we obtain
\begin{align}
{\cal A}^{(\pm)}_{\alpha\mu\bar\rho} %(t\rightarrow\infty)
=\sum_\nu\int^\infty_{-\infty}\frac{d\omega}{2\pi}\,
\Gamma^{(\pm)}_{\alpha\mu\nu}(\omega)
{\cal U}(\pm\omega)[a^{\pm}_\nu\bar\rho] .
\end{align}
Then, substituting this result into \Eq{SCBA-ME},
we can straightforwardly solve for $\bar\rho$
and calculate the steady state current.

We would like to mention that, remarkably,
for noninteracting system,
the steady state current given by this SCBA-ME scheme
coincides precisely with the nonequilibrium Green's function approach,
both giving the {\it exact} result \cite{LJL11}.
Notice also that, by contrast, the Born master equation
is applicable only to sequential tunneling
transport, being valid only in large bias limit.

%%  ================================================================

\section{Steady State Solution of the Anderson Impurity Model}

In \Eq{U-kondo}, associated with the steady state solution
of the Anderson impurity model, we have
\begin{align}
\lambda^+_\mu(\omega)&=i\frac{ \Pi^{-1}_{1\mu}(\omega)\bar\rho_{00}
-\Sigma^-_{\bar\mu}(\omega)\bar\rho_{\bar\mu\bar\mu}}
{ \Pi^{-1}_{\mu}(\omega)\Pi^{-1}_{1\mu}(\omega)},  \nonumber
\\
\lambda^-_\mu(\omega)&=i\frac{ \Pi^{-1}_{1\mu}(\omega)\bar\rho_{\mu\mu}
-\Sigma^-_{\bar\mu}(\omega)\bar\rho_{dd} }
{ \Pi^{-1}_{\mu}(\omega)\Pi^{-1}_{1\mu}(\omega)}, \nonumber
\\
\kappa^+_\mu(\omega)&=i\frac{ -\Sigma^+_{\bar\mu}(\omega)\bar\rho_{00}
+\Pi^{-1}_{\mu}(\omega)\bar\rho_{\bar\mu\bar\mu} }
{ \Pi^{-1}_{\mu}(\omega)\Pi^{-1}_{1\mu}(\omega)},  \nonumber
\\
\kappa^-_\mu(\omega)&=i\frac{ -\Sigma^+_{\bar\mu}(\omega)\bar\rho_{\mu\mu}
+\Pi^{-1}_{\mu}(\omega)\bar\rho_{dd} }
{ \Pi^{-1}_{\mu}(\omega)\Pi^{-1}_{1\mu}(\omega)}.   % \nonumber
\end{align}
%\esube
Here we introduced
$\Pi^{-1}_{ \mu}(\omega)=\omega-\epsilon_\mu-\Sigma_{0\mu}(\omega)
 -\Sigma^+_{\bar\mu}(\omega)$, and
$\Pi^{-1}_{1\mu}(\omega)=\omega-\epsilon_\mu-U-\Sigma_{0\mu}(\omega)
-\Sigma^-_{\bar\mu}(\omega)$.
The self-energy $\Sigma_{0\mu}(\omega)$ is given by
\begin{align}\label{Self1}
\Sigma_{0\mu}(\omega)
&=\int^\infty_{-\infty} \frac{d\omega'}{2\pi}
\frac{\Gamma_{\mu}(\omega')}{\omega-\omega' +i0^+} ,
\end{align}
while $\Sigma^{\pm}_{\mu}(\omega)$ by
 \begin{align}
\Sigma^{\pm}_{\mu}(\omega)&=\int \frac{d\omega'}{2\pi}
\frac{\Gamma^{(\pm)}_{\mu}(\omega')}{\omega-\epsilon_{\bar\mu}
+\epsilon_\mu-\omega' +i0^+}
\nl&
+\int \frac{d\omega'}{2\pi}
\frac{\Gamma^{(\pm)}_{\mu}(\omega')}{\omega-E_d+\omega' +i0^+}.
\end{align}
With the above results, as outlined after \Eq{U-kondo},
one is able to carry out the steady state solution $\bar{\rho}$.
Based on it, to obtain further the current, we first introduce
$\varphi_{1\mu\nu}(\omega)={\rm Tr} \big[a_\mu\ti\rho_{1\nu}(\omega)\big]$
and
$\varphi_{2\mu\nu}(\omega)={\rm Tr} \big[a_\mu\ti\rho_{2\nu}(\omega)\big]$,
where $\ti\rho_{1\nu}(\omega)$ and $\ti\rho_{2\nu}(\omega)$
are calculated using \Eq{tirhoj},
with an initial condition of
$\ti\rho_{1\nu}(0)=\bar\rho a^\dg_\nu$
and $\ti\rho_{2\nu}(0)=a^\dg_\nu\bar\rho$.
To simplify notations, we denote the various matrices
in boldface form: $\bm\varphi_1(\omega)$,
$\bm\varphi_2(\omega)$ and $\bm{\Gamma}_{L(R)}$.
Now, if $\bm\Gamma_L$ is proportional to $\bm\Gamma_R$ by a constant,
the steady state current can be recast to
the Landauer-B\"uttiker type, in terms of an integration of
tunneling coefficient over the incident energies,
$ \bar I = \frac{2e}{\hbar}{\rm Re}
\int^\infty_{-\infty}\frac{d\omega}{2\pi}
\left[ n_L(\omega)- n_R(\omega)\right] {\cal T}(\omega) $.
The tunneling coefficient, very compactly, is given by
%\begin{align}
$  {\cal T}(\omega)={\rm Tr}\{
  \bm\Gamma_L\bm\Gamma_R
  (\bm\Gamma_L+\bm\Gamma_R)^{-1}
  {\rm Re}\big[\bm\varphi(\omega)\big]\}$,
%\end{align}
where $\bm\varphi(\omega)=\bm\varphi_1(\omega)+\bm\varphi_2(\omega)$.
%%
%Compared to the nGF formulation \cite{Hau96}, we find that
%$\bm\varphi$ plays a role of the retarded Green's function,
%i.e., $\bm\varphi(\omega)=i\bm G^r(\omega)$.
%%
For the Anderson impurity system in nonequilibrium, we find
\begin{align}\label{Kondo-A}
& \bm\varphi(\omega)
=\frac{i\big[\Pi^{-1}_{1\mu}(\omega)
-\Sigma^{(+)}_{\bar\mu}(\omega)\big]
(1-n_{\bar\mu})}{\Pi^{-1}_{\mu}(\omega)\Pi^{-1}_{1\mu}(\omega)
-\Sigma^{(+)}_{\bar\mu}(\omega)\Sigma^{(-)}_{\bar\mu}(\omega)}  \nl
& ~~~~~~~~
+\frac{i\big[\Pi^{-1}_{\mu}(\omega)-\Sigma^{(-)}_{\bar\mu}(\omega)\big]
n_{\bar\mu}}{\Pi^{-1}_{\mu}(\omega)\Pi^{-1}_{1\mu}(\omega)
-\Sigma^{(+)}_{\bar\mu}(\omega)\Sigma^{(-)}_{\bar\mu}(\omega)}
 \nl
&=
\frac{i(1-n_{\bar\mu})}
{\omega-\epsilon_\mu-\Sigma_{0\mu}
 +U\Sigma^+_{\bar\mu}(\omega-\epsilon_\mu-U-\Sigma_{0\mu}
 -\Sigma_{\bar\mu})^{-1}
 }   \nl
%\nla
& ~~
+\frac{ i n_{\bar\mu}}
{\omega-\epsilon_\mu-U-\Sigma_{0\mu}
 -U\Sigma^-_{\bar\mu}(\omega-\epsilon_\mu-\Sigma_{0\mu}
 -\Sigma_{\bar\mu})^{-1}    },
\end{align}
where $n_\mu=\rho_{\mu\mu}+\rho_{dd}$, and
$1-n_\mu=\rho_{\bar\mu\bar\mu}+\rho_{00}$.
This result, precisely, coincides with that given by
the EOM technique of the nonequilibrium Green's function \cite{Hau96}.
One can check that, as discussed in detail in Ref.\ \cite{Hau96},
this solution contains the nonequilibrium Kondo effect.

\clearpage
%\begin{thebibliography}{99}

%\clearpage

%\begin{figure}\label{Fig3}
%\caption{}
%\end{figure}

%\end{CJK}
\end{document}